\documentclass[pre,twocolumn,showpacs,preprintnumbers,amsmath,amssymb]{revtex4}

\usepackage{graphicx}
\usepackage{bm}
\usepackage{xcolor}

\begin{document}
\title{Scaling law for the transient behavior of type-II neuron
models}

\author{M. A. D. Roa} 
\author{M. Copelli} \email{mcopelli@df.ufpe.br}
\affiliation{Laborat\'orio de F\'{\i}sica Te\'orica e Computacional,
Departamento de F\'{\i}sica, Universidade Federal de Pernambuco,
50670-901 Recife, PE, Brazil} \author{O. Kinouchi}
\affiliation{Faculdade de Filosofia,
Ci\^encias e Letras de Ribeir\~ao Preto, Universidade de S\~ao Paulo,
Avenida dos Bandeirantes 3900, 14040-901, Ribeir\~ao Preto, SP, Brazil}
\author{N. Caticha} 
\affiliation{Instituto de F\'{\i}sica, Universidade de S\~ao Paulo
05508-090, S\~ao Paulo, SP, Brazil}

\begin{abstract}
We study the transient regime of type-II biophysical neuron models and
determine the scaling behavior of relaxation times $\tau$ near but
below the repetitive firing critical current, $\tau \simeq C
(I_c-I)^{-\Delta}$. For both the Hodgkin-Huxley and Morris-Lecar
models we find that the critical exponent is independent of the
numerical integration time step and that both systems belong to the
same universality class, with $\Delta = 1/2$. For appropriately chosen
parameters, the FitzHugh-Nagumo model presents the same generic
transient behavior, but the critical region is significantly
smaller. We propose an experiment that may reveal nontrivial critical
exponents in the squid axon.

\end{abstract}


\maketitle

\section{Introduction}


The search for biophysical models for information processing systems
has led to a variety of model neurons which describe the dynamics of
the membrane potential. Collective properties arise from their
interaction through several possible architectures and types of
couplings. These can be chemical, voltage-gated synapses, simpler
proteic electrical connections, or even just electrical ephaptic
interactions arising between neighboring nerve fibers. Maybe the most
striking feature of a neuron is the threshold of the stimulus that
separates spiking from nonspiking regimes.  Spiking neurons generate
taletale signatures which have served as the basis for
frequency-dependent neural codes, an idea that can be traced back to
the work of Adrian in the 1920s~\cite{Adrian26}. Although of paramount
importance to neural dynamics, spike frequencies do not tell the
complete story. Subtle computations may arise from subthreshold
dynamics such as for examples in the early stages of the mammalian
visual system, olfactory bulb, and cortex.  In many cases the key to
the information dynamics lies in the transients, either below the
current threshold to generate action potentials or the threshold to
generate infinite sequences of spikes.  In this paper we investigate
transient spike trains of single model neurons since this might have a
bearing on the collective behavior, i.e., computational capabilities,
of subthreshold assemblies of neurons.

A dynamical system approach reveals that universal bifurcation
scenarios for the firing behavior appear irrespective of the specific
membrane ion channels and microscopic details involved. Full
characterization of these bifurcation routes is important for a deeper
understanding of how they affect the firing behavior and possible
implementation of neural codes. For example, it is now acknowledged
that bistable behavior and the small range of firing frequencies in
neurons that undergo subcritical Hopf bifurcations prevent a robust
use of a pure frequency code.

The transient behavior of neuron models has not received much
attention in either experimental or theoretical studies. In this
paper we study the divergence of transient times in a class of
conductance-based models and show that it follows a universal
critical behavior. We propose an experiment that may test our
theoretical predictions and discuss how neurons could employ
transients for computational purposes.

\section{Transients in type-II models}
The Hodgkin-Huxley (HH) model is a biophysically motivated system of
four coupled nonlinear differential equations that describe the
dynamics of the membrane potential $V$ of the squid giant axon
(e.g., \cite{Johnston94,Koch}):

\begin{eqnarray}
C\frac{dV}{dt} & = & G_{Na} m^3h\left(E_{Na}-V\right)+G_{K} n^4
\left(E_{K}-V\right) \nonumber \\
 && + G_L\left(E_{L}-V\right) + I(t), \nonumber \\
\frac{dx_i}{dt} & = & \alpha_{x_i}(V)(1-x_i)-\beta_{x_i}(V)x_i \; .
\label{eqHH}
\end{eqnarray}
$x_i$ stand for the three gate variables $x_i = m,h,$ and $n$
describing the activation of ionic channels and $\alpha_{x_i}$,
$\beta_{x_i}$ are voltage-dependent transition rates~\cite{Koch}:

\begin{eqnarray}
\alpha_m(V) & = & \frac{2.5-0.1V}{e^{(2.5-0.1V)} - 1}, \nonumber \\
\beta_m(V) & = & 4\, e^{-V/18}, \nonumber \\
\alpha_h(V) & = & 0.07\, e^{-V/20}, \nonumber \\
\beta_h(V) & = & \frac{1}{e^{(3-0.1V)} + 1}, \nonumber \\
\alpha_n(V) & = & \frac{0.1-0.01V}{e^{(1-V)} - 1}, \nonumber \\ 
\beta_n(V)& =& 0.125\, e^{-V/20} \; , 
\end{eqnarray}
where voltages are expressed in mV, rates in ms$^{-1}$,
$C=1$~$\mu$F/cm$^2$ is the specific membrane capacitance,
$G_{Na}=120$~mS/cm$^2$, $G_{K}=36$~mS/cm$^2$, and $G_{L}=0.3$~mS/cm$^2$
are ionic conductances per unit area, and $E_{Na}=115$~mV,
$E_{K}=-12$~mV, and $E_{L}=10.6$~mV are reversal potentials.

The HH system plays a fundamental role in the field of neurophysiology
and computational neuroscience since it defines the class of
conductance-based models.  As a function of a constant applied current
$I$, the HH model undergoes a subcritical Hopf bifurcation at $I=I_H$,
above which the fixed point solution (membrane potential at rest) is
no longer stable and trajectories are attracted to a stable limit
cycle, leading to repetitive firing (infinite train of action
potentials).

\begin{figure}[!tb]
\begin{center}
\includegraphics[width=1\columnwidth]{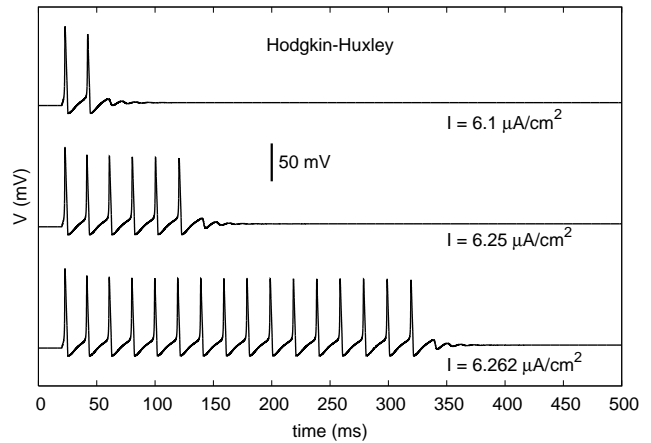}
\caption{\label{transHH} Examples of transient behavior near $I_c$ for
the HH model.  The constant step current is applied at $t=10$ ms. The
estimated critical current is $I_c= 6.26422125685$~$\mu$A/cm$^2$ for
an integration time step $dt=0.01$~ms.
}
\end{center}
\end{figure}

The sudden jump to a periodic behavior with nonzero frequency $f$ is
analogous to a first-order phase transition and is referred to as type-II
behavior in the neuroscience literature~\cite{Rinzel98b}. As in
first-order phase transitions, coexistence also appears in type-II
behavior. Just below the Hopf bifurcation, the stable fixed point
coexists with a stable limit cycle, their basins of attraction being
separated by an unstable limit cycle~\cite{Rinzel98b}. Both limit
cycles are created in a saddle-node (or ``fold'') bifurcation of cycles
at $I=I_c < I_H$. In our analogy with equilibrium phase transitions,
$I_c$ would correspond to a spinodal point. If the fixed point at
$I=0$ is perturbed by the application of a constant current near but
below $I_c$, several spikes may appear before the system returns to
the new resting state (Fig.~\ref{transHH}).

\begin{figure}[!bt]
\begin{center}
\includegraphics[width=1\columnwidth]{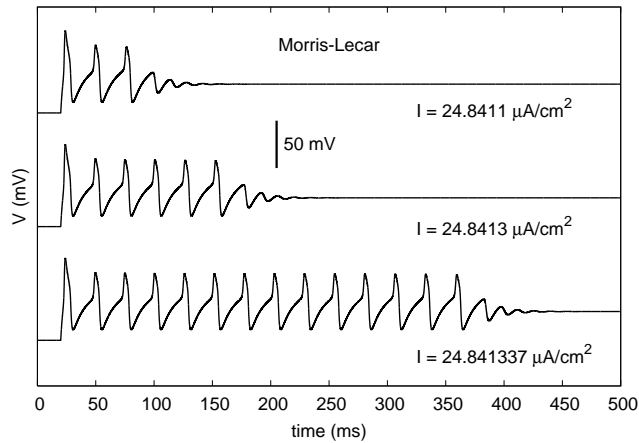}
\caption{\label{transML} Examples of transient behavior near $I_c$ for
the ML system. The constant step current is applied at $t=10$ ms. The
estimated critical current is $I_c=24.84134676279$~$\mu$A/cm$^2$ for
an integration time step $dt=0.01$ ms.}
\end{center}
\end{figure}

In the original version of the Morris-Lecar (ML)
model~\cite{Morris-Lecar}, a system with two coupled nonlinear ordinary differential equaions (ODEs)
used to describe action potentials in a barnacle motor fiber, the
relevant bifurcation at the onset of repetitive firing is a
saddle-node one.  That means that the spiking frequency varies
continuously from $I_c$ as $f \propto (I-I_c)^\beta$, with $\beta =
1/2$, which is similar to a mean field second-order phase transition
behavior if we think of $f$ as the order parameter. This transition is
called type-I behavior in the neuroscience literature~\cite{Rinzel98b}
and does not present a slow transient phenomenon similar to Fig.~1.
However, the Morris-Lecar system has also been used to describe cells
which present a type-II behavior~\cite{Rinzel98b,Koch}, which occurs
for the equations

\begin{eqnarray}
C\frac{dV}{dt} & = & 0.5 G_{Ca}
 \left[1+\tanh\left(\frac{V+1}{15}\right)\right] \left(E_{Ca}-V\right)
 \nonumber \\ && + G_{K}w\left(E_{K}-V\right) +
 G_L\left(E_{L}-V\right) + I(t), \nonumber \\ \frac{dw}{dt} & = &
 0.1\cosh(V/60)\left[1+\tanh\left(V/30\right)-2w\right]\; ,
\end{eqnarray}
when, for example, the values of the parameters are chosen as
$G_{Ca}=1.1$~mS/cm$^2$, $G_{K}=2.0$~mS/cm$^2$, $G_{L}=0.5$~mS/cm$^2$,
$E_{Ca}=100$~mV, $E_{K}=-70$~mV, and $E_{L}=-50$~mV. Then, large
transient times are also observed~(Fig.~\ref{transML}).

\begin{figure}[!bt]
\begin{center}
\includegraphics[width=1\columnwidth]{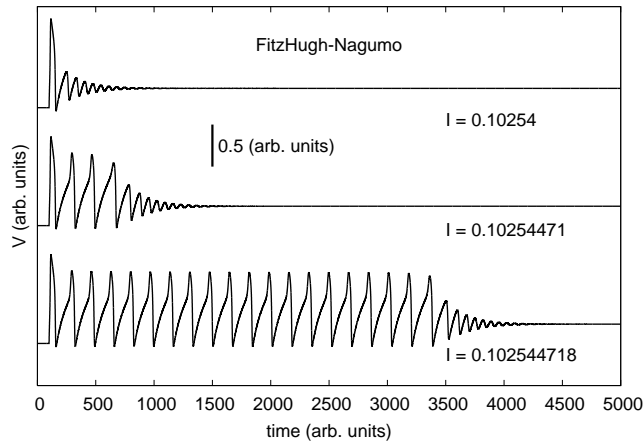}
\caption{\label{transFHN} Examples of transient behavior near $I_c$
for the FHN system. The constant step current is applied at
$t=10$. The estimated critical current is $I_c=0.1025447183127$ for an
integration time step $dt=0.01$ (all quantities in arbitrary units).}
\end{center}
\end{figure}

The FitzHugh-Nagumo (FHN) system

\begin{eqnarray}
\frac{dV}{dt} & = & V(V-a)(1-V) -w +I,  \nonumber \\
\frac{dw}{dt} & = & \epsilon (V-\gamma w)\; ,
\end{eqnarray}
has been proposed as a low-dimensional toy model that represents the
type-II behavior of the HH and other excitable systems. We verified
that the transient behavior here reported is not seen with the usual
parameters~\cite{Rinzel77}. However, the FHN model can reproduce the
HH transient behavior if one chooses parameters near $a=0.5$,
$\gamma=4.2$, and $\epsilon =0.01$ (Fig.~\ref{transFHN}). 


In this paper we show that the long relaxation times in type-II models
are a consequence of the changes in phase space which occur near the
creation of the limit cycles. Moreover, this leads to a scaling relation
whose exponent can be predicted, in agreement with numerical
simulations.

\section{Scaling law}

The relaxation time $\tau$ may be defined as the time until the last
spike, or the time until the membrane voltage stays within a small
distance from the resting potential (these two times are very similar
near $I_c$). When we plot $\tau$ as a function of $I_c-I$ we find a
power law divergence of the relaxation time, $\tau \simeq C
(I_c-I)^{-\Delta}$, where $\Delta$ is similar to a dynamic critical
exponent. The $\Delta$ exponent characterizes the ``critical slowing
down'' behavior near the bifurcation.  We expect that $\Delta$ is a
universal exponent but we are not aware that this exponent has been
measured for neuron models.  Here we report the measured exponents for
these three type-II biophysical neuron models, finding very good
agreement between them.

\begin{figure}[!tb]
\begin{center}
\includegraphics[width=1\columnwidth]{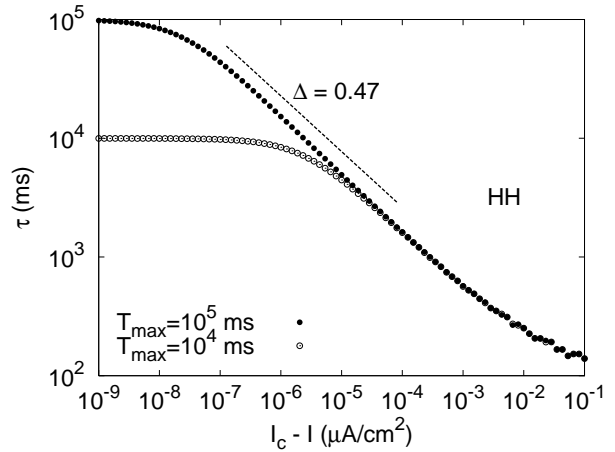}
\caption{\label{powHH} Relaxation times for the HH model as a function
of the distance to critical current for different integration times:
$T_{max}=10^4$~ms (open circles) and $T_{max}=10^5$~ms (filled
circles).}
\end{center}
\end{figure}

\begin{figure}[!tb]
\begin{center}
\includegraphics[width=1\columnwidth]{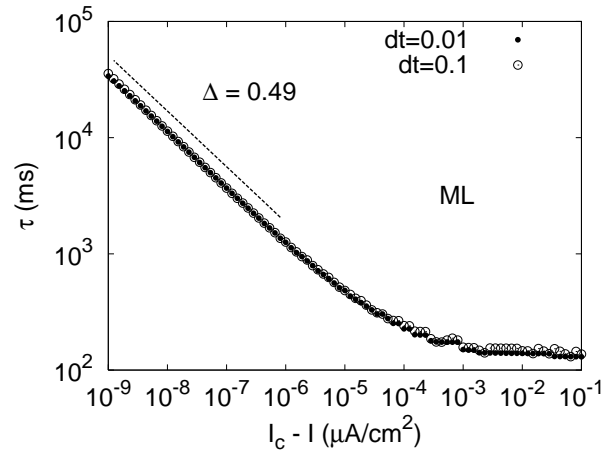}
\caption{\label{powML} Relaxation times for the Morris-Lecar model as a
function of the distance to critical current for different integration
times: $dt=0.01$ ms (filled circles) and $dt=0.1$ ms (open circles). }
\end{center}
\end{figure}

\begin{figure}[!tb]
\begin{center}
\includegraphics[width=1\columnwidth]{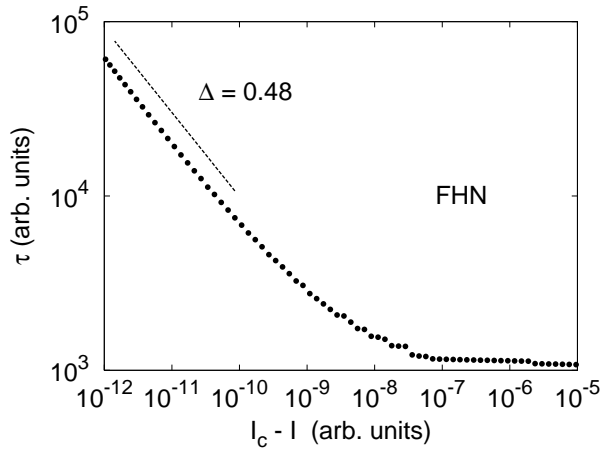}
\caption{\label{powFHN} Relaxation times for the FitzHugh-Nagumo model
as a function of the distance to critical current. Note that the power
law becomes visible only {\em very\/} close to the fold bifurcation
($I_c - I\sim 10^{-9}$).}
\end{center}
\end{figure}

We integrated the equations using a standard fourth-order Runge-Kutta
algorithm and determined $\tau$ by measuring the time interval from
the onset of the current step to a near stop of the flow [$|\dot{\bf
x}|< 10^{-5}$, where $\dot{\bf x}$ is the velocity vector in phase
space: $\dot{\bf x}= (\dot{w},\dot{v})$ for the ML and FHN systems,
and $\dot{\bf x}= (\dot{w},\dot{m},\dot{h},\dot{n})$ for the HH
model]. As opposed to the Hopf bifurcation, the fold bifurcation
cannot be obtained analytically, so $I_c$ was estimated numerically
after integration of the ODEs up to a (long) maximum time
$T_{max}$. The determination of the critical current is sensitive to
$T_{max}$, but in practice this only limits the range of validity of
the power law (see Fig.~\ref{powHH}). We have employed
$T_{max}=10^5$~ms and $dt=0.01$~ms, unless otherwise stated.  The
estimated critical currents $I_c(T_{max},dt)$ quoted in the figure
captions are very precisely determined given $T_{max}$ and the
integration step size $dt$.

The critical exponent is determined from the plot $\tau$ vs $I_c-I$.
We found $\Delta = 0.47$ for the HH system (Fig.~\ref{powHH}) and
$\Delta = 0.49$ for the ML model (Fig.~\ref{powML}), irrespective of
the size of the integration step $dt$.  This suggests a universal
exponent $\Delta = 1/2$. Obtaining long transients in the FHN model
has proved numerically more difficult, since the phenomenon occurs
only very close to $I_c$ (Fig.~\ref{powFHN}). Nonetheless, we have
obtained the exponent $\Delta = 0.48$.

\begin{figure}[!tb]
\begin{center}
\includegraphics[width=1\columnwidth]{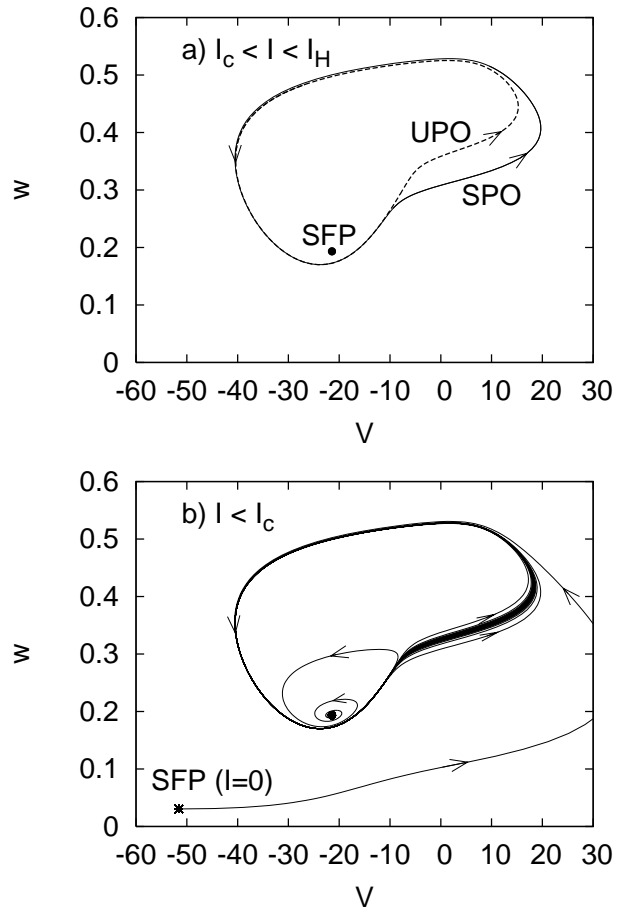}
\caption{\label{limitcycles} Phase portraits of the ML model for $I$
slightly above (a) and below (b) $I_c$. The long transient is
dominated by the time it takes to pass through the region where the
limit cycles are about to emerge. SPO (UPO) = Stable (Unstable)
Periodic Orbit, SFP = Stable Fixed Point. Horizontal axes (V) in mV and vertical axes (W) are dimensionless.}
\end{center}
\end{figure}

\begin{figure}[!bt]
\begin{center}
\includegraphics[width=1\columnwidth]{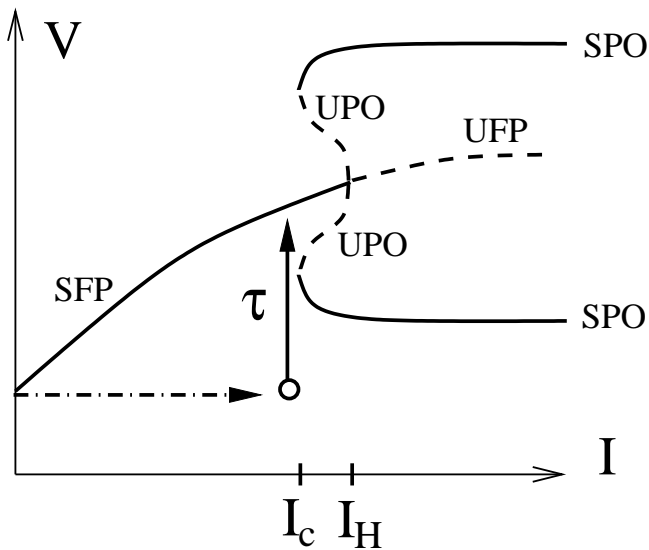}
\caption{\label{esquema} Schematic bifurcation diagram
for type II neuron models. The fold bifurcation occurs at $I_c$, while
the Hopf bifurcation occurs at $I_{H}$. Owing to the onset of the
current step (dot-dashed arrow), the fixed point for $I=0$ becomes the
initial condition in a new phase portrait. The transient $\tau$ (solid
arrow) to reach the new fixed point is governed by the ghost limit
cycle where the UPO and the SPO annihilate each other (see
Fig.~\ref{limitcycles}).}
\end{center}
\end{figure}

Figure~\ref{limitcycles}(a) shows that for $I_c \lesssim I < I_H$ an
unstable limit cycle and a stable one coexist and surround a stable fixed
point. The fixed point for $I=0$ lies outside both limit cycles
[Fig.~\ref{limitcycles}(b)]. Therefore, when the current is abruptly
changed to $I \lesssim I_c$, the fixed point is displaced to a region
within a ghost limit cycle, and the transient is completely
dominated by the time it takes for the system to overcome it. The
ghost is a natural consequence of the system being immediately
below a saddle-node bifurcation of cycles, and can be characterized by
the vanishingly small flow component normal to the half-stable limit
cycle that is created at $I=I_c$. It effectively works as a
one-dimensional extended bottleneck through which the system must pass
before reaching the fixed point (see Fig.~\ref{esquema} for a
caricature). If one considers an analogous system in polar
coordinates~\cite{Strogatz} $\dot\theta=f(r,\theta)$, $\dot r = \mu r
+ r^3 - r^5$, where $f(r,\theta)>0\;, \forall r>0$, it is clear that
for $\mu \lesssim \mu_c = -1/4$ the time for the system to overcome
the ghost at $r=1/2$ scales as $\tau \sim (\mu_c - \mu)^{-1/2}$
(the Hopf bifurcation occurring only at $\mu_H=0$). Solutions for the
transient behavior have been obtained by Tonnelier~\cite{Tonnelier02}
for McKean's piecewise linear version of the FHN
model~\cite{McKean70}.  However, the scaling behavior has not been
observed because the model has been studied in the absence of an
external current.

\begin{figure}[!tb]
\begin{center}
\includegraphics[width=1\columnwidth]{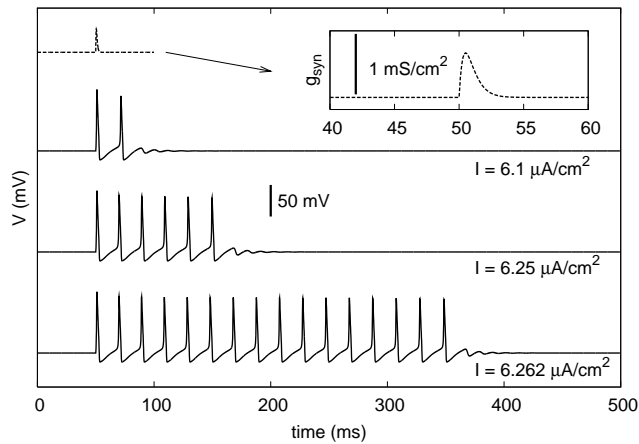}
\caption{\label{fig:epsp} Long transients appear if the system
initially at rest with $I\lesssim I_c$ is perturbed by an additional
EPSP (see text for details). Solid lines are membrane potentials,
while the dashed line is the synaptic conductance.}
\end{center}
\end{figure}

It should be clear that the current step is just a simple way of
putting the system outside the ghost limit cycle, but the scaling law
for the transient is not restricted to this somewhat artificial
protocol (even though it is very common in both theory and
experiments). Close to the fold transition, any short-lived
perturbation that is strong enough to make the system cross the ghost
limit cycle will give rise to long transients back to the fixed
point. We exemplify this with a biologically plausible example in the HH
model. Suppose the system is somehow maintained close to criticality
at $I\lesssim I_c$ (this could be achieved by several different
possible mechanisms, so we just fix $I$). In addition, assume the
neuron undergoes a fast excitatory postsynaptic potential (EPSP)
simulated by an injected synaptic current
$I_{syn}(t)=g_{syn}(t)(E_{Na}-V)$:

\begin{eqnarray}
C\frac{dV}{dt} & = & G_{Na} m^3h\left(E_{Na}-V\right)+G_{K} n^4
\left(E_{K}-V\right) \nonumber \\
&& + G_L\left(E_{L}-V\right) + I +I_{syn}(t)\; .
\end{eqnarray}
The fast change in the synaptic conductance is given by Rall's alpha
function~\cite{Wilson99}: $g(t^\prime) = \theta(t^\prime)(g_m
t^\prime/\tau_s^2 )\exp(-t^\prime/\tau_s)$, where $\theta$ is the
Heaviside function, $g_m=1$~mS/cm$^2$, $\tau_s=0.5$~ms, and
$t^\prime=t-t_{EPSP}$, where $t_{EPSP}=50$~ms is the time the EPSP is
initiated. Results are shown in Fig.~\ref{fig:epsp}, whose similarity
with Fig.~\ref{transHH} attests to the robustness of the effect
(notice however that, differently from Fig.~\ref{transHH}, in
Fig.~\ref{fig:epsp} the resting membrane potential is at the $I\neq 0$
fixed point {\em before\/} the perturbation). This opens interesting
possibilities from the point of view of neuronal computation: the
length of the transient response of a neuron ``probed'' by an EPSP
could code for its internal state of excitability, that is, for how
close it is to $I_c$. Naturally, this transient coding would work only
if the system is close to the fold bifurcation. We therefore
could have another example in neuroscience of optimal information
processing at criticality~\cite{Camalet00, Beggs03, Chialvo04,
Eguiluz05, Furtado06, Kinouchi06a, Chialvo06}.

It is interesting to point out that this bottleneck effect is
analogous to what occurs for type-I neurons above the
saddle-node bifurcation that leads to repetitive firing. In that case,
however, the ghost results from the anihilation of fixed points
(not limit cycles) and the period $T$ of the limit cycles
diverges as $(I-I_c)^{-1/2}$. This provides a complementary scenario
connecting both classes of neurons: the transient of type-II models
below $I_c$ diverges with the same exponent as the period of type-I
models above $I_c$, that is, $\Delta = \beta$.

\section{Concluding remarks}

We were unable to find a description of this scaling law behavior for
transients in neuron models or the associated dynamic critical
exponent in the literature.  Some kind of critical slowing down for
subthreshold oscillations has been reported experimentally in the
squid axon~\cite{Matsumoto78}, but these authors examined the vicinity
of a parametric subcritical Hopf bifurcation, not a saddle-node
bifurcation of cycles induced by external currents.  We propose that a
similar experiment with high-precision injected currents near $I_c$
could be used to check the power law found in the computational
model. Since the area of the giant squid axon is of order $\sim
1$~cm$^2$, to examine the critical regime requires that current
fluctuations should be less than $\sim 10^3$~pA (see
Fig.~\ref{powHH}). Even if the full critical regime seems to be hard
to achieve, the initial divergence in the transient lifetime may
provide an experimental check of our predictions.

We emphasize that both in standard experiments and in our single-compartment
model space clamping is used. It might happen though, as
in spin systems, that for the extended real system without voltage
clamp, or for a compartmental model with a large number of compartments,
the exponents may differ from the values here reported, changing the
universality class to one not described by a ``mean field''
approach. Interestingly, this would mean that collective properties
within a nontrivial universality class could be observed at the level
of a single axon.  So whether this result can be verified
experimentally hinges on the effects that spatially extended neurons
may have on the robustness of this picture. 

Furthermore, noise could always play a role. In studies of type-I
intermittency in simple maps, the length of the ``laminar phase''
$\left< l\right>$ in a chaotic regime is analogous to the transient in
this work and diverges as $\left< l\right> \sim \epsilon^{-1/2}$
because of a zero-dimensional bottleneck as the distance $\epsilon$ to
a tangent bifurcation tends to
zero~\cite{Manneville79,Manneville80}. In the presence of additive
noise with amplitude $g$~\cite{Hirsch82}, the scaling changes to
$\sqrt{\epsilon}\left< l\right> \sim f(g^2/\epsilon^{3/2})$, where $f$
is a universal function (see also~\cite{Cavalcante04}
for recent extensions). It is conceivable that similar scaling
relations could be obtained for $\tau$ provided that the chaotic phase
of intermittency theory could be replaced by some mechanism of
``reinjection'' in type-II neuron models.  For instance, in the
phenomenon of coherence resonance~\cite{coherenceresonance} noise
itself plays this reinjecting role.  However, the excitable systems
employed are usually not close enough to the fold transition to
exhibit long transients, so it would be interesting to investigate the
effects of the scaling laws we report here in the resonance curves.
These theoretical issues should be dealt with before engaging in an
experimental search.

{\textbf Acknowledgments:} The research was supported by FACEPE, CNPq and
special program PRONEX. M.A.D.R. thanks
the organizers of LASCON, the Latin American School on Computational Neuroscience,
for financial support. M.C. thanks
J. R. Rios Leite and J. R. L. de Almeida for inspiring discussions.
\bibliography{copelli}

\end{document}